\documentclass{acm_proc_article-sp}
\usepackage{graphicx, verbatim, color}

\begin{document}

\title{PageRank: Standing on the shoulders of giants}

\numberofauthors{1}
\author{
\alignauthor
Massimo Franceschet \\
\affaddr{Department of Mathematics and Computer Science} \\
\affaddr{University of Udine} \\
\affaddr{Via delle Scienze 206, 33100 Udine, Italy} \\
\email{massimo.franceschet@uniud.it}}

\maketitle

\section{Introduction}

PageRank~\cite{BP98} is a Web page ranking technique that has been a fundamental ingredient in the development and success of the Google search engine.  The method is still one of the many signals that Google uses to determine which pages are most important.\footnote{\texttt{http://www.google.com/corporate/tech.html}}  The main idea behind PageRank is to determine the importance of a Web page in terms of the importance assigned to the pages hyperlinking to it. In fact, this thesis is not new, and has been previously successfully exploited in different contexts. We review the PageRank method and link it to some renowned previous techniques that we have found in the fields of Web information retrieval, bibliometrics, sociometry, and econometrics.

\section{Web information retrieval} \label{web}

In 1945 Vannevar Bush wrote a today celebrated article in \textit{The Atlantic Monthly} entitled ``As We May Think'' describing a futuristic device he called Memex~\cite{B45}. Bush writes:

\begin{quote}
\textit{``Wholly new forms of encyclopedias will appear, ready made with a \emph{mesh} of associative trails running through them, ready to be dropped into the Memex and there amplified.''}
\end{quote}

Bush's prediction came true in 1989, when Tim Berners-Lee proposed the \textit{Hypertext Markup Language} (HTML) to keep track of experimental data at the European Organization for Nuclear Research (CERN). In the original far-sighted proposal in which Berners-Lee attempts to persuade CERN management to adopt the new global hypertext system we can read the following paragraph\footnote{\texttt{http://www.w3.org/History/1989/proposal.html}}:

\begin{quote}
\textit{``We should work toward a universal linked information system, in which generality and portability are more important than fancy graphics techniques and complex extra facilities. The aim would be to allow a place to be found for any information or reference which one felt was important, and a way of finding it afterwards. The result should be sufficiently attractive to use that the information contained would grow past a critical threshold.''}
\end{quote}

As we all know, the proposal was accepted and later implemented in a \textit{mesh} -- this was the only name that Berners-Lee originally used to describe the Web -- of interconnected documents that rapidly grew beyond the CERN threshold, as Berners-Lee anticipated, and became the World Wide Web.

Today, the Web is a huge, dynamic, self-organized, and hyperlinked data source, very different from traditional document collections which are nonlinked, mostly static, centrally collected and organized by specialists. These features make Web information retrieval quite different from traditional information retrieval and call for new search abilities, like automatic crawling and indexing of the Web. {\color{black} Moreover, early search engines ranked responses using only a \textit{content score}, which measures the similarity between the page and the query.  One simple example is just a count of the number of times the query words occur on the page, or perhaps a weighted count with more weight on title words.} These traditional query-dependent techniques suffered under the gigantic size of the Web and the death grip of spammers.

In 1998, Sergey Brin and Larry Page revolutionised the field of Web information retrieval by introducing the notion of an \textit{importance score}, which gauges the status of a page, independently from the user query, by analysing the topology of the Web graph. The method was implemented in the famous PageRank algorithm and both the traditional content score and the new importance score were efficiently combined in a new search engine named Google.

\section{Ranking Web pages using PageRank} \label{pagerank}

We briefly recall how the PageRank method works keeping the mathematical machinery to the minimum.
Interested readers can more thoroughly investigate the topic in a recent book of Langville and Meyer which elegantly describes the science of search engine rankings in a rigorous yet playful style~\cite{LM06}.

We start by providing an intuitive interpretation of PageRank in terms of random walks on graphs~\cite{PBMW99}. The Web is viewed as a directed graph of pages connected by hyperlinks. A \textit{random surfer} starts from an arbitrary page and simply keeps clicking on successive links at random, bouncing from page to page. The PageRank value of a page corresponds to the relative frequency that the random surfer visits that page, assuming that the surfer goes on infinitely. The more time spent by the random surfer on a page, the higher the PageRank importance of the page.

A little more formally, the method can be described as follows. Let us denote by $q_i$ the number of distinct outgoing (hyper)links of page $i$. Let $H = (h_{i,j})$ be a square matrix of size  equal to the number $n$ of Web pages such that $h_{i,j} = 1 / q_i$ if there exists a link from page $i$ to page $j$ and $h_{i,j} = 0$ otherwise. The value $h_{i,j}$ can be interpreted as the probability that the random surfer moves from page $i$ to page $j$ by clicking on one of the distinct links of page $i$. The PageRank $\pi_j$ of page $j$ is \textit{recursively} defined as: $$\pi_j = \sum_i \pi_i h_{i,j}$$ or, in matrix notation, $\pi = \pi H$.
Hence, the PageRank of page $j$ is the sum of the PageRank scores of pages $i$ linking to $j$, weighted by the probability of going from $i$ to $j$. In words, the PageRank thesis reads as follows:
\begin{quote}
\textit{A Web page is important if it is pointed to by other important pages.}
\end{quote}
There are in fact three distinct factors that determine the PageRank of a page: (i) the number of links it receives, (ii) the link propensity, that is, the number of outgoing links, of the linking pages, and (iii) the PageRank of the linking pages. The first factor is not surprising: the more links a page receives, the more important it is perceived. Reasonably, the link value depreciates proportionally to the number of links given out by a page: endorsements coming from parsimonious pages are worthier than those emanated by spendthrift ones. Finally, not all pages are created equal: links from important pages are more valuable than those from obscure ones.

Unfortunately, this ideal model has two problems that prevent the solution of the system. The first one is due to the presence of \textit{dangling nodes}, that are pages with no forward links.\footnote{The term \textit{dangling} refers to the fact that many dangling nodes are in fact pendent Web pages found by the crawling spiders but whose links have not been yet explored.}  These pages capture the random surfer indefinitely. Notice that a dangling node corresponds to a row in matrix $H$ with all entries equal to 0. {\color{black} To tackle the problem of dangling nodes, the corresponding rows in $H$ are replaced by the uniform probability vector $u = 1/n \, e$, where $e$ is a vector of length $n$ with all components equal to 1. Alternatively, one may use any fixed probability vector in place of $u$.} This means that the random surfer escapes from the dangling page by jumping to a randomly chosen page. We call $S$  the resulting matrix.

The second problem with the ideal model is that the surfer can get trapped into a \textit{bucket} of the Web graph, which is a reachable strongly connected component without outgoing edges towards the rest of the graph. The solution proposed by Brin and Page is to replace matrix $S$ by the \textit{Google matrix} $$G = \alpha S + (1-\alpha) E$$ where $E$ is the \textit{teleportation matrix} with identical rows each equal to the uniform probability vector $u$, and $\alpha$ is a free parameter of the algorithm often called the \textit{damping factor}. Alternatively, a fixed \textit{personalization} probability vector $v$ can be used in place on $u$. In particular, the personalization  vector can be exploited to bias the result of the method towards certain topics.
The interpretation of the new system is that, with probability $\alpha$ the random surfer moves forward by following links, and, with the complementary probability $1 - \alpha$ the surfer gets bored of following links and enters a new destination in the browser's URL line, possibly unrelated to the current page. The surfer is hence teleported, like a Star Trek character, to that page, even if there exists no link connecting the current and the destination pages in the Web universe.  The inventors of PageRank propose to set the damping factor $\alpha = 0.85$, meaning that after about five link clicks the random surfer chooses a random page.

The PageRank vector is then defined as the solution of equation:

\begin{equation} \label{PR2} \pi = \pi G \end{equation}

\begin{figure}[t]
\begin{center}
\includegraphics[scale=0.50, angle=0]{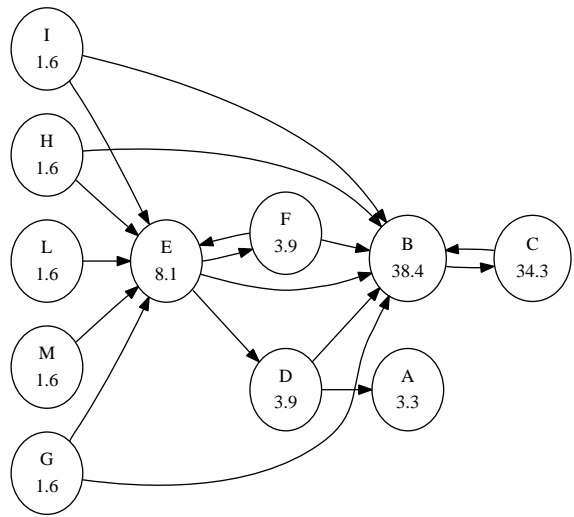}
\caption{A PageRank instance with solution.  Each node is labelled with its PageRank score. Scores have been normalized to sum to 100.  We assumed $\alpha = 0.85$.}
\label{pagerank-ex}
\end{center}
\end{figure}

An example is provided in Figure~\ref{pagerank-ex}. Node A is a dangling node, while nodes B and C form a bucket. Notice the dynamics of the method: page C receives just one link but from the most important page B; its importance is much higher than that of page E, which receives many more links, but from anonymous pages. Pages G, H, I, L, and M do not receive endorsements; their scores correspond to the minimum amount of status of each page.

Typically, the normalization condition $\sum_i \pi_i = 1$ is also added. In this case Equation~\ref{PR2} becomes $\pi = \alpha \pi S + (1-\alpha) u$. The latter distinguishes two factors contributing to the PageRank vector: an \textit{endogenous} factor equal to $\pi S$ which takes into consideration the real topology of the Web graph, and an \textit{exogenous} factor equal to the uniform probability vector $u$, which can be interpreted as a minimal amount of status assigned to each page independently of the hyperlink graph. The parameter $\alpha$ balances between these two factors.

\section{Computing the PageRank vector} \label{computation}

Does Equation~\ref{PR2} have a solution? Is the solution unique? Can we efficiently compute it? The success of the PageRank method rests on the answers to these queries. Luckily, all these questions have nice answers.

Thanks to the dangling nodes patch, matrix $S$ is a stochastic matrix\footnote{This simply means that all rows sum up to 1.}, and clearly the teleportation matrix $E$ is also stochastic. It follows that $G$ is stochastic as well, since it is defined as a convex combination of stochastic matrices $S$ and $E$. It is easy to show that, if $G$ is stochastic, Equation~\ref{PR2} has always at least one solution. Hence, we have got at least one PageRank vector. Having two independent PageRank vectors, however, would be already too much: which one should we use to rank Web pages? Here, a fundamental result of algebra comes to the rescue : \textit{Perron-Frobenius theorem}~\cite{P07,F12}. It states that, if $A$ is an irreducible\footnote{A matrix is irreducible if and only if the directed graph associated with it is strongly connected, that is, for every pair $i$ and $j$ of graph nodes there are paths leading from $i$ to $j$ and from $j$ to $i$.} nonnegative square matrix, then there exists a unique vector $x$, called the Perron vector, such that $x A = rx$, $x > 0$, and $\sum_i x_i = 1$, where $r$ is the maximum eigenvalue of $A$ in absolute value, that algebraists call the \textit{spectral radius}  of $A$. The Perron vector is the left \textit{dominant eigenvector} of $A$, that is, the left eigenvector associated with the largest eigenvalue in magnitude.

The matrix $S$ is most likely reducible, since experiments have shown that the Web has a bow-tie  structure fragmented into four main continents that are not mutually reachable, as first observed in~\cite{BKMRRSTW00}. Thanks to the teleportation trick, however, the graph of matrix $G$ is strongly connected. Hence $G$ is irreducible and Perron-Frobenius theorem applies\footnote{Since $G$ is stochastic, its spectral radius is 1.}. Therefore, a positive PageRank vector exists and is furthermore unique.

Interestingly, we can arrive at the same result using \textit{Markov theory}~\cite{M06}. The above described random walk on the Web graph, modified with the teleportation jumps, naturally induces a finite-state \textit{Markov chain}, whose transition matrix is the stochastic matrix $G$. Since $G$ is irreducible, the chain has a unique \textit{stationary distribution} corresponding to the PageRank vector.

A last crucial question remains: can we efficiently compute the PageRank vector? The success of PageRank is largely due to the existence of a fast method to compute its values: the \textit{power method}, a simple iteration method to find the dominant eigenpair of a matrix developed by von Mises and Pollaczek-Geiringer~\cite{MP29}. It works as follows on the Google matrix $G$. Let $\pi^{(0)} = u = 1/n \, e$. Repeatedly compute  $\pi^{(k+1)} = \pi^{(k)} G$ until $||\pi^{(k+1)} - \pi^{(k)}|| < \epsilon$, where $||\cdot||$ measures the distance between the two successive PageRank vectors and $\epsilon$ is the desired precision.

The convergence rate of the power method is approximately the rate at which $\alpha^k$ approaches to $0$: the closer $\alpha$ to unity, the lower the convergence speed of the power method. If, for instance, $\alpha = 0.85$, as many as 43 iterations are sufficient to gain 3 digits of accuracy, and 142 iterations are enough for 10 digits of accuracy. Notice that the power method applied to matrix $G$ can be easily expressed in terms of matrix $H$, which, unlike $G$, is a very sparse matrix that can be stored using a linear amount of memory with respect to the size of the Web.

\section{Standing on the shoulders of giants}

\begin{table}[t]
\begin{center}
{\color{black}
\begin{tabular}{lll}
\hline
\textbf{Year} &	\textbf{Author} & \textbf{Contribution} \\ \hline
1906 & Markov & Markov theory~\cite{M06} \\
1907 & Perron & Perron theorem~\cite{P07} \\
1912 & Frobenius & Perron-Frobenius theorem~\cite{F12} \\
1929 & von Mises \& & Power method~\cite{MP29}  \\
     & Pollaczek-Geiringer & \\
1941 & Leontief & Econometric model~\cite{L41} \\
1949 & Seeley & Sociometric model~\cite{S49} \\
1952 & Wei & Sport ranking model~\cite{W52} \\
1953 & Katz & Sociometric model~\cite{K53} \\
1965 & Hubbell & Sociometric model~\cite{H65} \\
1976 & Pinski \& Narin & Bibliometric model~\cite{PN76} \\
1998 & Kleinberg & HITS~\cite{K98} \\
1998 & Brin \& Page & PageRank~\cite{BP98} \\ \hline
\end{tabular}
}
\end{center}
\caption{PageRank history.}
\label{history}
\end{table}

\textit{Dwarfs standing on the shoulders of giants} is a Western metaphor meaning ``One who develops future intellectual pursuits by understanding the research and works created by notable thinkers of the past''.\footnote{From the Wikipedia page for \textit{Standing on the shoulders of giants}.} The metaphor was famously uttered by Isaac Newton: \textit{``If I have seen a little further it is by standing on the shoulders of Giants''}. Moreover, \textit{``Stand on the shoulders of giants''} is Google Scholar's motto: \textit{``the phrase is our acknowledgement that much of scholarly research involves building on what others have already discovered''}.

{\color{black} There are many giants upon whose shoulders PageRank firmly stands: Markov~\cite{M06}, Perron~\cite{P07}, Frobenius~\cite{F12}, von Mises and Pollaczek-Geiringer~\cite{MP29} provided at the beginning of the 1900's the necessary mathematical machinery to investigate and effectively solve the PageRank problem.} Moreover, the circular PageRank thesis has been previously exploited in different contexts, including Web information retrieval, bibliometrics, sociometry, and econometrics. In the following, we review these contributions and link them to the PageRank method. {\color{black} Table~\ref{history} contains a brief summary of PageRank history.} All the ranking techniques surveyed in this paper have been implemented in R~\cite{R07} and the code is freely available at the author's Web page.

\subsection{Hubs and authorities on the Web}

\begin{figure}[t]
\begin{center}
\includegraphics[scale=0.50, angle=0]{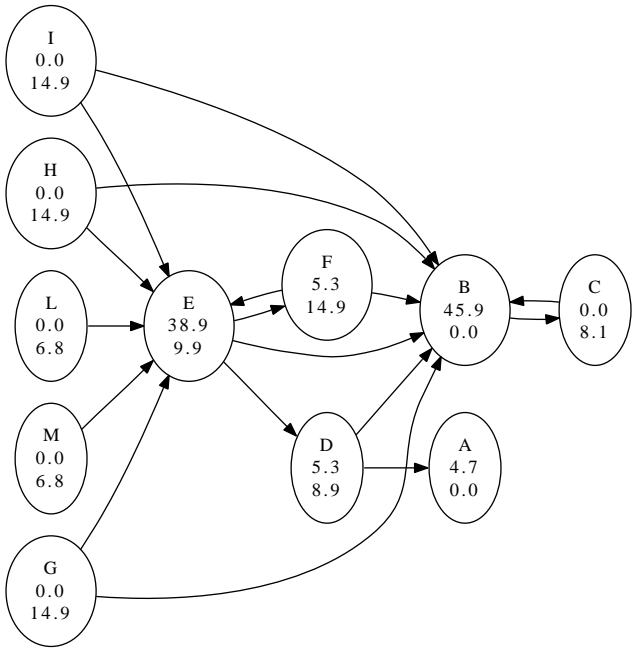}
\caption{A HITS instance with solution (compare with PageRank scores in Figure~\ref{pagerank-ex}). Each node is labelled with its authority (top) and hub (bottom) scores. Scores have been normalized to sum to 100. The dominant eigenvalue for both authority and hub matrices is 10.7.}
\label{hits-ex}
\end{center}
\end{figure}

\textit{Hypertext Induced Topic Search} (HITS) is a Web page ranking method proposed by Jon Kleinberg~\cite{K98, K99}. The connections between HITS and PageRank are striking. Despite the close conceptual, temporal and even geographical proximity of the two approaches, it appears that HITS and PageRank have been developed independently. In fact, both papers presenting PageRank~\cite{BP98} and HITS~\cite{K99} are today citational blockbusters: the PageRank article collected 6167 citations, while the HITS paper has been cited 4617 times.\footnote{Source: Google Scholar on February 5, 2010.}

HITS thinks of Web pages as \textit{authorities} and \textit{hubs}. HITS circular thesis reads as follows:

\begin{quote}
\textit{Good authorities are pages that are pointed to by good hubs and good hubs are pages that point to good authorities.}
\end{quote}

Let $L = (l_{i,j})$ be the adjacency matrix of the Web graph, i.e., $l_{i,j} = 1$ if page $i$ links to page $j$ and $l_{i,j} = 0$ otherwise. We denote with $L^{T}$ the transpose of $L$. HITS defines a pair of recursive equations as follows, where $x$ is the authority vector containing the authority scores and $y$ is the hub vector containing the hub scores:
\begin{equation} \label{hits1}
\begin{array}{lcl}
x^{(k)} & = & L^{T} y^{(k-1)} \\
y^{(k)} & = & L x^{(k)}
\end{array}
\end{equation}

where $k \geq 1$ and $y^{(0)} = e$, the vector of all ones. The first equation tells us that authoritative pages are those pointed to by good hub pages, while the second equation claims that good hubs are pages that point to authoritative pages. Notice that Equation~\ref{hits1} is equivalent to:

\begin{equation} \label{hits2}
\begin{array}{lcl}
x^{(k)} & = & L^{T} L x^{(k-1)} \\
y^{(k)} & = & L L^{T} y^{(k-1)}
\end{array}
\end{equation}

It follows that the authority vector $x$ is the dominant right eigenvector of the authority matrix $A = L^{T} L$, and the hub vector $y$ is the dominant right eigenvector of the hub matrix $H = L L^{T}$. This is very similar to the PageRank method, except the use of the authority and hub matrices instead of the Google matrix.

To compute the dominant eigenpair (eigenvector and eigenvalue) of the authority matrix we can again exploit the power method as follows: let $x^{(0)} = e$. Repeatedly compute  $x^{(k)} = A x^{(k-1)}$ and normalize $x^{(k)} = x^{(k)} / m(x^{(k)})$, where $m(x^{(k)})$ is the signed component of maximal magnitude, until the desired precision is achieved. It follows that $x^{(k)}$ converges to the dominant eigenvector $x$ (the authority vector) and $m(x^{(k)})$ converges to the dominant eigenvalue (the spectral radius, which is not necessarily 1). The hub vector $y$ is then given by $y = L x$. While the convergence of the power method is guaranteed, the computed solution is not necessarily unique, since the authority and hub matrices are not necessarily irreducible. {\color{black} A modification similar to the teleportation trick used for the PageRank method can be applied to HITS to recover the uniqueness of the solution~\cite{ZNJ01}.}

An example of HITS is given in Figure~\ref{hits-ex}. We stress the difference among importance, as computed by PageRank, and authority and hubness, as computed by HITS. Page B is both important and authoritative, but it is not a good hub. Page C is important but by no means authoritative. Pages G, H, I are neither  important nor authoritative, but they are the best hubs of the network, since they point to good authorities only. Notice that the hub score of B is 0 although B has one outgoing edge; unfortunately for B, the only page C linked by B has no authority. Similarly, C has no authority because it is pointed to only by B, whose hub score is zero. This shows the difference between indegree and authority, as well as between outdegree and hubness. {\color{black} Finally, we observe that  nodes with null authority  scores (respectively, null hub scores) correspond to isolated nodes in the graph whose adjacency matrix is the authority matrix $A$ (respectively, the hub matrix $H$).}

An advantage of HITS with respect to PageRank is that it provides two scores at the price of one. The user is hence provided with two rankings: the most authoritative pages about the research topic, which can be exploited to investigate in depth a research subject, and the most hubby pages, which correspond to portal pages linking to the research topic from which a broad search can be started. A disadvantage of HITS is the higher susceptibility of the method to spamming: while it is difficult to add incoming links to our favourite page, the addition of outgoing links is much easier. This leads to the possibility of purposely inflating the hub score of a page, indirectly influencing also the authority scores of the pointed pages.

{\color{black} An following algorithm that incorporates ideas from both PageRank and HITS is SALSA~\cite{LM00}: like HITS, SALSA computes both authority and hub scores, and like PageRank, these scores are obtained from Markov chains.}

\subsection{Bibliometrics} \label{sec:PN}

Bibliometrics, also known as scientometrics, is the quantitative study of the process of scholarly publication of research achievements. The most mundane aspect of this branch of information and library science is the design and application of \textit{bibliometric indicators} to determine the influence of bibliometric units like scholars and academic journals. The Impact Factor is, undoubtedly, the most popular and controversial journal bibliometric indicator available at the moment. It is defined, for a given journal and a fixed year, as the mean number of citations in the year to papers published in the two previous years. It has been proposed in 1963 by Eugene Garfield, the founder of the Institute for Scientific Information (ISI), working together with Irv Sher~\cite{GS63}. Journal Impact Factors are currently published in the popular Journal Citation Reports by Thomson-Reuters, the new owner of the ISI.

The Impact Factor  does not take into account the importance of the citing journals: citations from highly reputed journals are weighted as those from obscure journals. In 1976 Gabriel Pinski and Francis Narin developed an innovative journal ranking method~\cite{PN76}. The method measures the influence of a journal in terms of the influence of the citing journals. The Pinski and Narin thesis is:

\begin{quote}
\textit{A journal is influential if it is cited by other influential journals.}
\end{quote}

This is the same circular thesis of the PageRank method. Given a source time window $T_1$ and a previous target time window $T_2$, the journal citation system can be viewed as a \textit{weighted} directed graph in which nodes are journals and there is an edge from journal $i$ to journal $j$ if there is some article published in $i$ during $T_1$ that cites an article published in $j$ during $T_2$. The edge is weighted with the number $c_{i,j}$ of such citations from $i$ to $j$.  Let $c_i = \sum_j c_{i,j}$ be the total number of cited references of journal $i$.

In the method described by Pinski and Narin, a citation matrix $H = (h_{i,j})$ is constructed such that $h_{i,j} = c_{i,j}/ {c_j}$. The coefficient $h_{i,j}$ is the amount of citations received by journal $j$ from journal $i$ per reference given out by journal $j$. For each journal an  \textit{influence score} is determined which measures the relative journal performance per given reference. The influence score $\pi_j$ of journal $j$ is defined as: $$\pi_j = \sum_i \pi_i \frac{c_{i,j}}{c_j} = \sum_i \pi_i h_{i,j}$$ or, in matrix notation:

\begin{equation} \label{PN} \pi = \pi \, H \end{equation}

Hence, journals $j$ with a large \textit{total influence} $\pi_j c_j$ are those that receive significant endorsements from influential journals. Notice that the influence per reference score $\pi_j$ of a journal $j$ is a size independent measure, since the formula normalizes by the number of cited references $c_j$ contained in articles of the journal, which is an estimation of the size of the journal. Moreover, the normalization neutralizes the effect of journal self-citations, that are citations between articles in the same journal. These citations are indeed counted both at the numerator and at the denominator of the influence score formula. This avoids over inflating journals that engage in the practice of opportunistic self-citations.

{\color{black} It can be proved that the spectral radius of matrix $H$ is 1, hence the influence score vector corresponds to the dominant eigenvector of $H$~\cite{G78}.} In principle, the uniqueness of the solution and the convergence of the power method to it are not guaranteed. Nevertheless, both properties are not difficult to obtain in real cases. If the citation graph is strongly connected, then the solution is unique. When journals belong to the same research field, this condition is typically satisfied. Moreover, if there exists a self-loop in the graph, that is an article that cites an article in the same journal, then the power method converges.

\begin{figure}[t]
\begin{center}
\includegraphics[scale=0.5, angle=0]{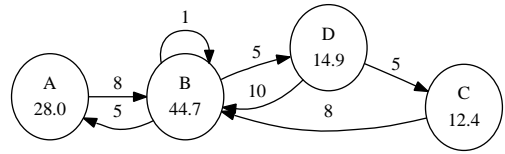}
\caption{An instance with solution of the journal ranking method proposed by Pinski and Narin. Nodes are labelled with influence scores and edges with the citation flow between journals. Scores have been normalized to sum to 100.}
\label{journalrank-ex}
\end{center}
\end{figure}

Figure~\ref{journalrank-ex} provides an example of the Pinski and Narin method. Notice that the graph is strongly connected and has a self-loop, hence the solution is unique and can be computed with the power method. Both journals A and C receive the same number of citations and give out the same number of references. Nevertheless, the influence of A is bigger, since it is cited by a more influential journal (B instead of D). Furthermore, A and D receive the same number of citations  from the same journals, but D is larger than A, since it contains more references, hence the influence of A is higher.

Similar recursive methods have been independently proposed by~\cite{LP84} and~\cite{PV04} in the context of ranking of economics journals. Recently, various PageRank-inspired bibliometric indicators to evaluate the importance of journals using the academic citation network have been proposed and extensively tested: journal PageRank~\cite{BRS06}, Eigenfactor~\cite{Eigenfactor07}, and SCImago~\cite{SCImago07}.

\subsection{Sociometry}

Sociometry, the quantitative study of social relationships, contains remarkably old PageRank predecessors. Sociologists were the first to use the network approach to investigate the properties of groups of people related in some way. They devised measures like indegree, closeness, betweeness, as well as eigenvector centrality which are still used today in modern (not necessarily social) network analysis~\cite{N10}. In particular,  eigenvector centrality uses the same central ingredient of PageRank applied to a social network:

\begin{quote}
\textit{A person is prestigious if he is endorsed by prestigious people.}
\end{quote}

John R. Seeley in 1949 is probably the first in this context to use the circular argument of PageRank~\cite{S49}. Seeley reasons in terms of social relationships among children: each child chooses other children in a social group with a nonnegative strength. The author notices that the total choice strengths received by each children is inadequate as an index of popularity, since it does not consider the popularity of the chooser. Hence, he proposes to define the popularity of a child as a function of the popularity of those children who chose the child, and the popularity of the choosers as a function of the popularity of those who chose them and so in an ``indefinitely repeated reflection''. Seeley exposes the problem in terms of linear equations and uses Cramer's rule to solve the linear system. He does not discuss the issue of uniqueness.

Another model is proposed in 1953 by Leo Katz~\cite{K53}. Katz views a social network as a directed graph where nodes are people and person $i$ is connected by an edge to person $j$ if $i$ chooses, or endorses, $j$. The status of member $i$ is defined as the number of weighted paths reaching $j$ in the network, a generalization of the indegree measure. Long paths are weighted less than short ones, since endorsements devalue over long chains. Notice that this method indirectly takes account of who endorses as well as how many endorse an individual: if a node $i$ points to a node $j$ and $i$ is reached by many paths, then the paths leading to $i$ arrive also at $j$ in one additional step.

Katz builds an adjacency matrix $L = (l_{i,j})$ such that $l_{i,j} = 1$ if person $i$ chooses person $j$ and $l_{i,j} = 0$ otherwise. He defines a matrix $W = \sum_{k=1}^{\infty} (a L)^k$, where $a$ is an attenuation constant. Notice that the $(i,j)$ component of $L^k$ is the number of paths of length $k$ from $i$ to $j$, and this number is attenuated by $a^k$ in the computation of $W$. Hence, the $(i,j)$ component of the limit matrix $W$ is the weighted number of arbitrary paths from $i$ to $j$. Finally, the status of member $j$ is $\pi_j = \sum_i w_{i,j}$, that is, the number of weighted paths reaching $j$. If the attenuation factor $a  < 1 / \rho(L)$, with $\rho(L)$ the spectral radius of $L$, then the above series for $W$ converges.

\begin{figure}[t]
\begin{center}
\includegraphics[scale=0.50, angle=0]{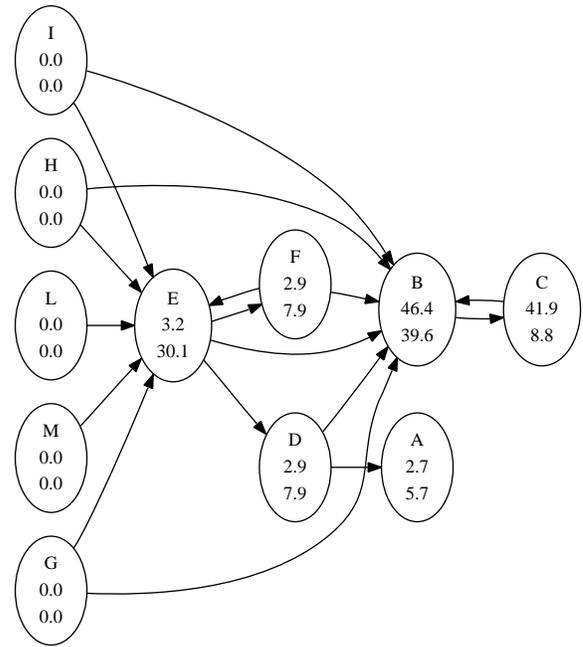}
\caption{An example of the Katz model using two attenuation factors: $a = 0.9$ and $a = 0.1$ (the spectral radius of the adjacency matrix $L$ is 1). Each node is labelled with the Katz score corresponding to $a = 0.9$ (top) and $a = 0.1$ (bottom). Scores have been normalized to sum to 100.}
\label{katz-ex}
\end{center}
\end{figure}

Figure~\ref{katz-ex} illustrates the method with an example. Notice the important role of the attenuation factor: when it is large (close to $1 / \rho(L)$), long paths are devalued smoothly, and Katz scores are strongly correlated with PageRank ones. In the shown example, PageRank and Katz methods provide the same ranking of nodes when the attenuation factor is 0.9. On the other hand, if the attenuation factor is small (close to 0), then the contribution given by paths longer than 1 rapidly declines, and thus Katz scores converge to indegrees, the number of incoming links of nodes. In the example, when the attenuation factor drops to 0.1, nodes C and E switch their positions in the ranking: node E, which receives many short paths, significantly increases its score, while node C, which is the destination of just one short path and many (devalued) long ones, significantly decreases its score.

In 1965 Charles H. Hubbell generalizes the proposal of Katz~\cite{H65}. Given a set of members of a social context, Hubbell defines a matrix $W = (w_{i,j})$ such that $w_{i,j}$ is the strength at which $i$ endorses $j$. Interestingly, these weights can be arbitrary, and in particular, they can be negative.  The prestige of a member is recursively defined in terms of the prestige of the endorsers and takes account of the endorsement strengths:

\begin{equation} \label{Hu} \pi = \pi \, W + v \end{equation}

The term $v$ is an \textit{exogenous} vector such that $v_i$ is a minimal amount of status assigned to $i$ from outside the system.

The original aspects of the method are the presence of an exogenous initial input and  the possibility of giving negative endorsements. A consequence of negative endorsements is that the status of an actor can also be negative. An actor that receives a positive (respectively, negative) judgment from a member of positive status increases (respectively, decreases) his prestige. On the other hand, and interestingly, receiving a positive judgment from a member of negative status makes a negative contribution to the prestige of the endorsed member (if you are endorsed by some person affiliated to the Mafia your reputation might drop indeed). Moreover, receiving a negative endorsement from a member of negative status makes a positive contribution to the prestige of the endorsed person (if the same Mafioso opposes you, then your reputation might raise).

\begin{figure}[t]
\begin{center}
\includegraphics[scale=0.5, angle=0]{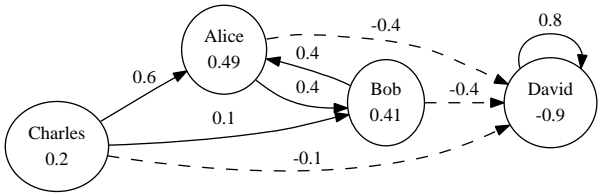}
\caption{An instance of the Hubbell model with solution: each node is labelled with its prestige score and each edge is labelled with the endorsement strength between the connected members; negative strength is highlighted with dashed edges. The minimal amount of status has been fixed to 0.2 for all members.}
\label{fig:hubbell}
\end{center}
\end{figure}

Figure~\ref{fig:hubbell} shows an example for the Hubbell model. Notice that Charles does not receive any endorsement and hence has the minimal amount of status given by default to each member. David receives only negative judgments; interestingly, the fact that he has a positive self opinion further decreases his status. A better strategy for him, knowing in advance of his negative status, would be to negatively judge himself, acknowledging the negative judgment given by the other members.

Equation~\ref{Hu} is equivalent to $\pi (I - W) = v$, where $I$ is the identity matrix, that is $\pi = v (I-W)^{-1} = v \, \sum_{i=0}^{\infty} W^i$. The series converge if and only if the spectral radius of $W$ is less than 1. It is now clear that the Hubbell model is a generalization of the Katz model to general matrices that adds an initial exogenous input $v$. Indeed, Katz equation for social status is $\pi = e \, \sum_{i=1}^{\infty} (a L)^i$, where $e$ is a vector of all ones. In an unpublished note Vigna traces the history of the mathematics of spectral ranking and shows that there is a reduction from the path summation formulation of Hubbell-Katz to the eigenvector formulation with teleportation of PageRank and vice versa~\cite{V10}. In the mapping the attenuation constant is the counterpart of the PageRank damping factor, and the exogenous  vector corresponds to the PageRank personalization vector. {\color{black} The interpretation of PageRank as a sum of weighted paths is also investigated in~\cite{BBC07}.}

{\color{black} Spectral ranking methods have been also exploited to rank sport teams in competitions that involve teams playing in pairs~\cite{W52,K55}. The underlying idea is that a team is strong if it won against other strong teams. Much of the art of the sport ranking problem is how to define the matrix entries $a_{i,j}$ expressing how much team $i$ is better than team $j$ (e.g., we could pick $a_{i,j}$ to be 1 if $j$ beats $i$, 0.5 if the game ended in a tie, and 0 otherwise)~\cite{K93}.}

\subsection{Econometrics}

We conclude with a succinct description of the input-output model developed in 1941 by Nobel Prize winner Wassily W. Leontief in the field of econometrics -- the quantitative study of economic principles~\cite{L41}. According to the Leontief  input-output model, the economy of a country may be divided into any desired number of sectors, called industries, each consisting of firms producing a similar product. Each industry requires certain inputs in order to produce a unit of its own product, and sells its products to other industries to meet their ingredient requirements. The aim is to find  prices for the unit of product produced by each industry that guarantee the \textit{reproducibility} of the economy, which holds when each sector balances the costs for its inputs  with the revenues of its outputs. In 1973, Leontief earned the Nobel Prize in economics for his work on the input-output model. An example is provided in Table~\ref{io}.

\begin{table*}[t]
\begin{center}
\begin{tabular*}{1\textwidth}{@{\extracolsep{\fill}}lrrr|rrr}
\hline  &	\textbf{agriculture} &	\textbf{industry} &	 \textbf{family} & \textbf{total} & \textbf{price} & \textbf{revenue} \\ \hline
\textbf{agriculture} & 7.5 & 6   & 16.5 & 30  & 20 & 600 \\ \hline
\textbf{industry}    & 14  & 6   & 30   & 50  & 15 & 750 \\ \hline
\textbf{family}      & 80  & 180 & 40   & 300 & 3  & 900\\ \hline \hline
\textbf{cost}        & 600 & 750 & 900 &&&\\ \hline
\end{tabular*}
\end{center}
\caption{An input-output table for an economy with three sectors with the balance solution. Each row shows the output of a sector to other sectors of the economy. Each column shows the inputs received by a sector from other sectors. For each sector we also show total quantity produced,  equilibrium unitary price,  total cost, and total revenue. Notice that each sector balances costs and revenues.}
\label{io}
\end{table*}

Let $q_{i,j}$ denote the quantity produced by the $i$th industry and used by the $j$th industry, and $q_i$ be the total quantity produced by sector $i$, that is, $q_i = \sum_j q_{i,j}$. Let $A = (a_{i,j})$ be such that $a_{i,j} = q_{i,j} / q_j$; each coefficient $a_{i,j}$ represents the amount of product (produced by industry) $i$ consumed by industry $j$ that is necessary to produce a unit of product $j$. Let $\pi_j$ be the price for the unit of product produced by each industry $j$. The reproducibility of the economy holds when each sector $j$ balances the costs for its inputs  with the revenues of its outputs, that is:
$$\begin{array}{lclc}
\mathit{cost}_j & = & \sum_i \pi_i q_{i,j}   = \\
\mathit{revenue}_j & = & \sum_i \pi_j q_{j,i}  = \pi_j \sum_i q_{j,i} = \pi_j q_j
\end{array}$$
By dividing each balance equation by $q_j$ we have $$\pi_j = \sum_i \pi_i \frac{q_{i,j}}{q_j} = \sum_i \pi_i a_{i,j} $$ or, in matrix notation, \begin{equation} \label{Leo} \pi = \pi A \end{equation}

Hence, highly remunerated industries (industries $j$ with high total revenue $\pi_j q_j$) are those that receive substantial inputs from highly remunerated industries, a circularity that closely resembles the PageRank thesis~\cite{PSS05}.  {\color{black} With the same argument used in~\cite{G78} for the Pinski and Narin bibliometric model we can show that the spectral radius of matrix $A$ is 1, thus the equilibrium price vector $\pi$ is the dominant eigenvector of matrix $A$.} Such a solution always exists, although it might not be unique, unless $A$ is irreducible.  Notice the striking similarity of the Leontief \textit{closed} model with that proposed by Pinski and Narin. An \textit{open} Leontief model adds an exogenous demand and creates a surplus of revenue (profit). It is described by the equation $\pi = \pi A + v$ where $v$ is the profit vector. Hubbell himself observes the similarity between his model and the Leontief open model~\cite{H65}.

It might seem disputable to juxtapose PageRank and Leontief methods. To be sure, the original motivation of Leontief work was to give a formal method to find equilibrium prices for the reproducibility of the economy and to use the method to estimate the impact on the entire economy of the change in demand in any sectors of the economy. Leontief, to the best of our limited knowledge, was not motivated by an industry \textit{ranking} problem. On the other hand, the motivation underlying the other methods described in this paper is the ranking of a set of homogeneous entities. Despite the original motivations, however, there are more than coincidental similarities between the Leontief open and closed models and the other ranking methods described in this paper. These connections motivated the discussion of the Leontief contribution, which is probably the least known among the surveyed methods within the computing community.

\section{Conclusion}

The classic notion of quality of information is related to the judgment given by few field \textit{experts}.  PageRank introduced an original notion of quality of information found on the Web: the \textit{collective intelligence} of the Web, formed by the opinions of the millions of people that populate this universe, is exploited to determine the importance, and ultimately the quality, of that information.

Consider the difference between \textit{expert evaluation} and \textit{collective evaluation}. The former tends to be intrinsic, subjective, deep, slow and expensive. By contrast, the latter is typically extrinsic, democratic, superficial, fast and low-cost. Interestingly, the dichotomy between these two evaluation methodologies is not peculiar to information found on the Web. In the context of assessment of academic research, peer review -- the evaluation of scholar publications given by peer experts working in the same field of the publication -- plays the role of expert evaluation. Collective evaluation consists in gauging the importance of a contribution though the bibliometric practice of counting and analysing citations received by the publication from the academic community. Citations generally witness the use of information and acknowledge intellectual debt. Eigenfactor~\cite{Eigenfactor07}, a PageRank-inspired bibliometric indicator, is among the most interesting recent proposals to collectively evaluate the status of academic journals. The consequences of a shift from peer review to bibliometric evaluation are currently heartily debated in the academic community~\cite{W05}.

\medskip
\noindent
\textbf{Acknowledgements}

\noindent The author thanks Enrico Bozzo, Sebastiano Vigna, and the anonymous referees for positive and critical comments on the early drafts of this paper. Sebastiano Vigna was the first to point out the contribution of John R. Seeley.

\bibliographystyle{abbrv}

\end{document}